\documentclass[fleqn,usenatbib]{rasti}

\usepackage{newtxtext,newtxmath}
\usepackage[T1]{fontenc}
\DeclareRobustCommand{\VAN}[3]{#2}
\let\VANthebibliography\thebibliography
\def\thebibliography{\DeclareRobustCommand{\VAN}[3]{##3}\VANthebibliography}
\usepackage{graphicx}	% Including figure files
\usepackage{amsmath}	% Advanced maths commands

\title[Limits to dynamic range in GHz-THz single-dish planetary spectra]{Limits to dynamic range in GHz-THz single-dish planetary spectra}

\author[J. S. Greaves]{
J. S. Greaves$^{1}$\thanks{E-mail: greavesj1@cardiff.ac.uk} 
\\
$^{1}$CHART, School of Physics \& Astronomy, Cardiff University, 4 The Parade, Cardiff CF24 3AA, UK
}

\date{Accepted XXX. Received YYY; in original form ZZZ}

\pubyear{2024}

\begin{document}
\label{firstpage}
\pagerange{\pageref{firstpage}--\pageref{lastpage}}
\maketitle

%It should be a single paragraph not more than 250 words.
\begin{abstract}

When bright solar-system objects are observed by GHz-THz regime telescopes, off-axis signals bounce around locally and re-enter the signal path with a time delay, causing sinusoidal ripples in output spectra. Ripples that are unstable over time are challenging to remove. A typical detection limit for planetary spectral lines is a fraction $\sim 10^{-3}$ of continuum signal, restricting searches for minor atmospheric trace-gases. Modern wideband spectra of Venus demonstrate a plethora of effects, at three example telescopes spanning nearly a factor-of-50 in frequency. Characterisation of instrumental effects as families of pure sine-waves via Fourier-analysis is shown to improve dynamic-range by factors of a few. An example upper limit on sulphuric acid (H$_2$SO$_4$) vapour in Venus' mesosphere, from fully-automated data-cleaning of a 3.5 GHz band containing 10 line components, goes as deep as the best previously-published limit. The most challenging cases are searches for single lines of width comparable to ripple periods. Traditional polynomial-fitting approaches can be deployed to test for false positives, to demonstrate robustness at a level of zero "fake lines" in >1000 comparisons. Fourier-based data-cleaning avoids subjectivity and can be fully automated, and synthetic spectra can be injected before processing to test to what degree signals are lost in cleaning. An ideal robustness strategy is mitigation at the data-acquisition stage, e.g. using slow drifts in target-velocity with respect to the telescope to isolate a planetary line from a quasi-static instrumental ripple pattern. 

\end{abstract}

% Include between one and six keywords.

%%%%%%%%%%%%%%%%% BODY OF PAPER %%%%%%%%%%%%%%%%%%

\section{Introduction}

\subsection{Context}

Spectra of solar system objects in centimetre to far-infrared bands can be very informative about their atmospheres. These wavelengths cover transitions between well-populated rotational energy-levels for many molecules, and can probe deep into thick atmospheres. In contrast, spectra at mid-infrared and shorter wavelengths tend to be of vibrational transitions, involving energy levels that may not be well populated in cooler atmospheres. These spectra tend also to be crowded with broad features, making minor trace gases hard to find. However, these wavebands are well suited to constructing compact instruments that can fly on spacecraft and so observe solar system bodies in-situ. For example, Mahieux et al. (2023) model observations at the day/night terminator made by the SOIR instrument on Venus Express, setting cloud-top upper limits for trace gases that are well below a part-per-billion (ppb). 

Planetary spectra at long wavelengths have decades of history (see e.g. de Pater \& Massie 1985), but are coming into more prominence as larger-scale observatories are built. Examples at the low/high frequency limits include the 35-50 GHz band of the ALMA interferometer and the 0.5-2 THz HIFI instrument that flew on ESA's Herschel. Modern wideband spectroscopy can offer up to $2^{14}$ spectral channels over $>$1 GHz bands, for spectral resolution as high as ${\rm d\nu/\nu} \sim 10^7$. The combination of high spectral resolution with wide bands is ideal for studying planetary atmospheres with a large range of pressures, as the line widths reflect pressure-broadening (collisional) processes. 

However, the sensitivity of such observations tends to be limited by instrumental effects, particularly where optical designs are not ideal for finding weak spectral features in the presence of a very strong planetary continuum. Molecules are seen in absorption against the planetary disc (in emission at the limb), and so a dynamic range can be defined as the detectable fraction of the continuum, or as the inverse of this value. Specialised observing techniques are needed to reach below a fraction of 10$^{-3}$. One example (Matthews et al. 2002) is the search for isotopically-substituted HCN in Jupiter's atmosphere after the SL-9 impact, needing nearly ten hours with the 15m JCMT to detect a fractional signal of $5 \times 10^{-4}$. Additional problems arise when the expected line width is poorly known, e.g. from a molecule at uncertain altitude, and thus pressure, within the atmosphere. This uncertainty can increase the chance of a false-positive arising from within a diverse set of instrumental artefacts.

\subsection{Instrumental issues}

Typically the presence of the planet's strong continuum signal introduces unwanted waves or "ripples" into the spectrum. In the GHz-THz regime, spectra are formed electronically, by feeding signals from the receiver (frontend) to a backend such as a correlator, that reconstructs intensity at each frequency using a Fourier transform. As described by e.g. Barnes, Briggs \& Calabretta (2005), off-axis signals can be reflected at surfaces around the telescope, re-entering the signal path with a time-delay. The Fourier transform of this time-delayed spike produces a sinusoidal ripple in the output spectrum, at a frequency of $c/2D$ where $D$ is the distance from the instrument to the reflecting surface. This extra component is not small when off-axis signals are from a bright planet. Ripples can also be generated by electronic reflections, i.e. standing waves between mismatched components.

Lack of control of such reflections is the main reason why dynamic range has not advanced greatly since Matthews et al. (2002) observed Jupiter over 25 years ago. Some strategies that have been developed include $\lambda/8$ defocussing at the Effelsberg 100m (Henkel 2005) and matching the ripple in the off-beam at the Mopra 22m (https://www.narrabri.atnf.csiro.au/mopra/obsinfo.html). In general, ripples that are stable over time are less problematic; for example, they could be characterised to good precision on a line-free target. Ripples that are not stable present a severe challenge, as it can be difficult to understand how they vary with e.g. telescope temperature, source elevation and stretching of cabling. 

The environments around telescopes are usually complex, making reflections difficult to suppress. For example, at large $D$, the support legs of a telescope's secondary mirror are reflective, and it would be challenging to clad these large structures in microwave-absorbent materials. Short-period ripples from these reflections hinder the detection of narrow lines from upper atmospheres. For thick atmospheres (producing more pressure-broadened wider lines), long ripple periods are a problem, from  strong reflections in the vicinity of the instrument. These effects can be very difficult to control due to a complicated environment with surfaces that may flex as the telescope moves. More fundamentally, the optical path will change as the receiver observes a calibration load, which is necessary to convert from raw signal counts to units of flux or antenna temperature. The difference in ripples between the on- and off-source spectra can produce very complex patterns in the calibrated spectra. Finally, non-periodic features may be evident in spectra, seen for example from resonances in cabling at the 15m JCMT -- a recent approach to removal of artefacts due to electronics is given by Liu et al. (2022), as applied at the FAST 500m\footnote{Telescope acronyms used in this work: ALMA = Atacama Large Millimeter/submillimeter Array; FAST = Five-hundred-meter Aperture Spherical Telescope; GBT = Green Bank Telescope; JCMT = James Clerk Maxwell Telescope; SKA = Square Kilometre Array; SOFIA = Stratospheric Observatory for Infrared Astronomy.}. 

The above problems apply to single-dish telescopes. For interferometers, reflections etc. arising at each antenna should not correlate, and so will not directly affect the interferometric output. However, final spectra can be affected by artefacts at individual antennas and on baselines between antennas; see Starr (2024) for an analysis for ALMA. In general, phase and amplitude errors can affect data fidelity (Richards et al. 2022). For observations of very extended sources such as planets, inadequate characterisation of the primary beam can also prevent reaching a good dynamic range (e.g. Greaves et al. 2021, for Venus). This can make it very difficult to achieve science goals, with for example 8/10 ALMA programs on Venus that have been completed since 2011 yet to be published. 

\begin{figure}
	\includegraphics[width=\columnwidth]{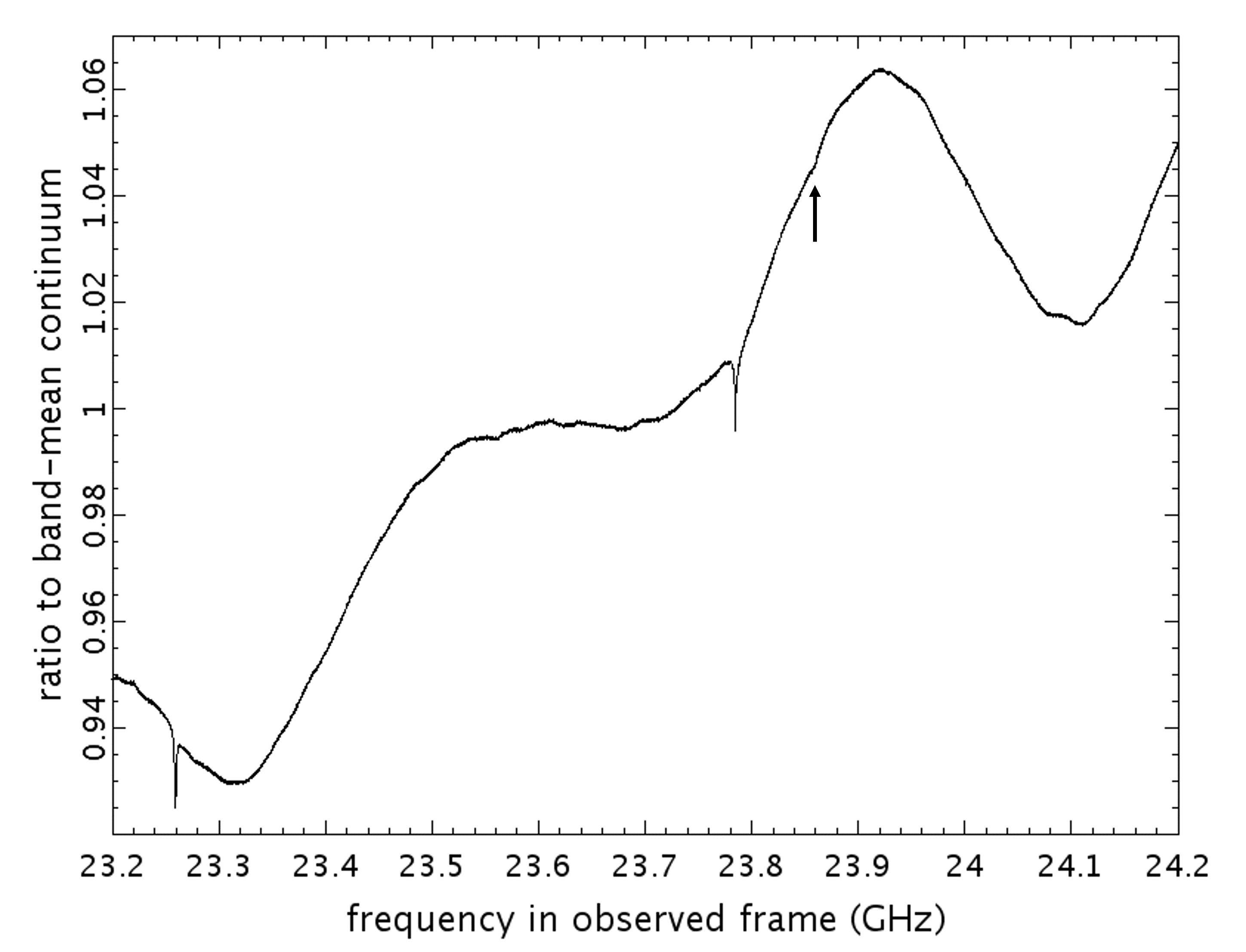}
    \caption{Selected data from GBT observations of Venus on 7 Feb 2022. {\it OnOff} observations with the KFPA frontend were averaged, and a 1 GHz section from the VEGAS backend is shown, for one of the two polarizations. The signals were normalised to the mean signal of the band shown. "V-shaped" dips are identified as instrumental, as they were not seen in other pixel/polarization combinations. The feature at rest frequency 23.8596 (terrestrial O$_3$), indicated by the arrow, is confirmed as real as it is seen in these other combinations.}
    \label{fig:1}
\end{figure}

\begin{figure}
    \centering
    \includegraphics[width=\columnwidth]{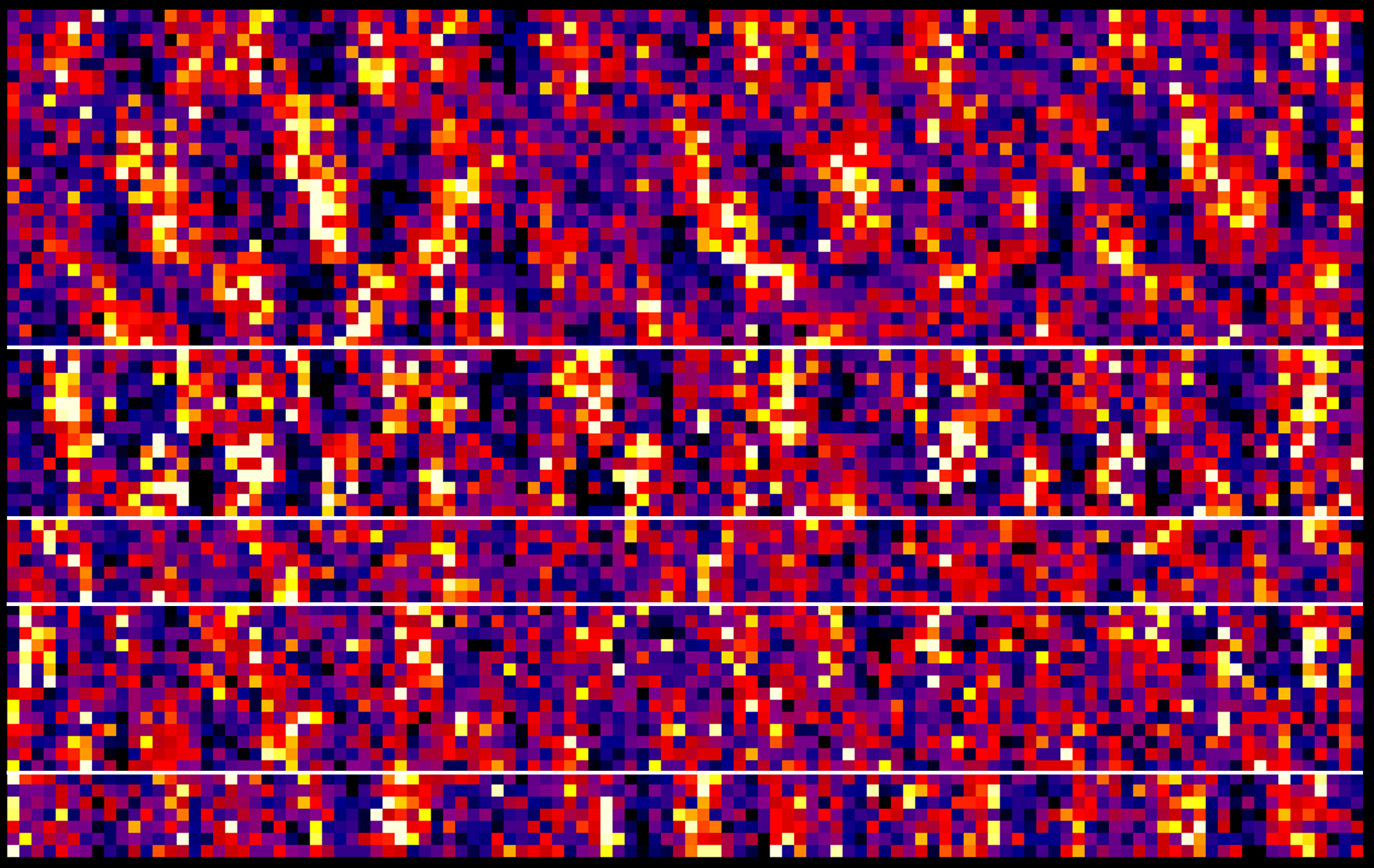}
    \includegraphics[width=\columnwidth]{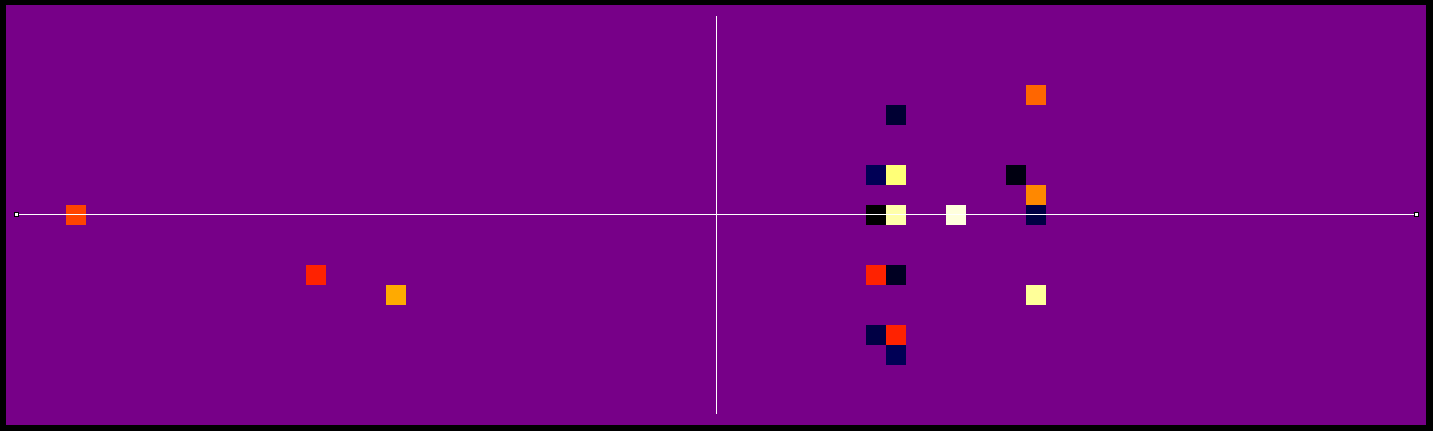}
    \caption{Top: the 2017 discovery dataset from JCMT for PH$_3$ J=1-0 at Venus (Greaves et al. 2021), displayed on axes of observation number (y) versus spectral channel (x) centred around 266.945 GHz. The original dataset is compressed here by factors of (64,2) in (x,y) for clarity and the linear colour-scale shows 95\% of the signal range. The white lines divide data acquired on 9, 11, 12, 14 and 16 June 2017, plotted from bottom to top. Bottom: same dataset displayed as the Hermitian component after a Fourier transform, with pixels $<8\sigma$ reset to be close to zero (purple). Lowest spatial frequencies occur where the white lines intersect. For scale, the narrowest ripple has 32 periods across the band (left-most red pixel) and 7 days data are included (period in e.g. orange pixel towards upper-right).  }
    \label{fig:1}
\end{figure}

\begin{figure*}
	\includegraphics[width=2.1\columnwidth]{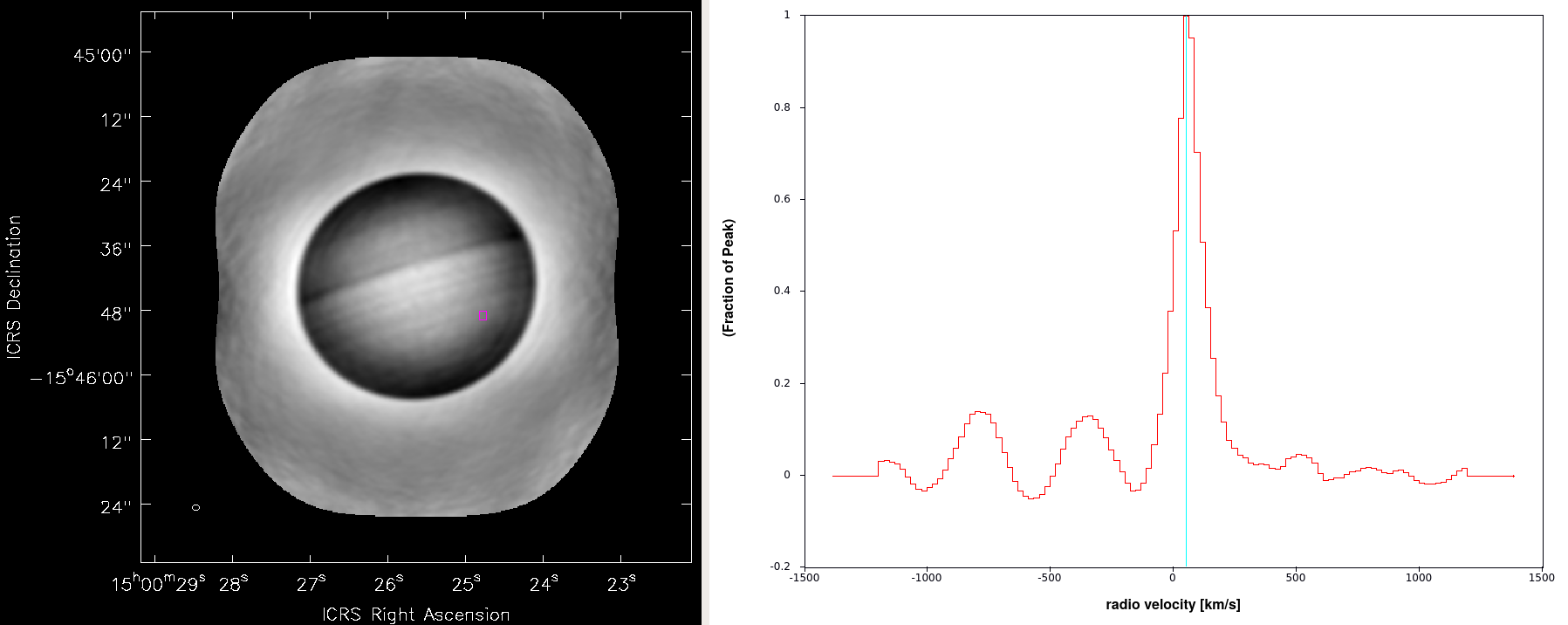}
    \caption{Example spectrum (at right) from the point marked (pink square in left panel) on the disc of Jupiter, as observed with ALMA on 16 May 2018. The datacube slice shown in the image is at the velocity shown by the blue vertical line in the spectrum. The spectrum is normalised to its peak line signal (not to the continuum signal as elsewhere in this work; no continuum data product was available). Data product was from the ALMA science archive, released 14 June 2019 for project id 2017.1.00279.S (PI I. de Pater). Images are from  the CASA viewer tool. }
    \label{fig:1}
\end{figure*}

\subsection{Aims}

This work aims to illustrate and quantify some dynamic range issues, primarily for single dish observations made with wide bandwidths. Some standard ripple-cleaning methods are assessed, to see what improvements to dynamic range can be made, and the need for new approaches is examined. The example data are from observations of Venus, comparing spectra at 24, 267 and 1067 GHz, acquired respectively with the 100m GBT, 15m JCMT and 2.5m SOFIA telescopes during 2021-2023. Single-beam spectra were acquired in each case, with Venus subtending an angular diameter similar to the full-width at half-maximum of the telescope beam.

\section{Characterisation}

\subsection{Dynamic range example}

An example single-dish spectrum of Venus is shown in Figure 1. This is a case where periodic ripples are minimal, because the GBT has an off-axis secondary mirror, reducing unwanted reflections. However, instrumental effects still produce an undulating spectral trend, that was not fully removed when a nearby off-source spectrum was subtracted. A real feature in this spectrum is the barely-visible dip from terrestrial ozone (rest frequency: 23.8596 GHz) that absorbs against the Venus continuum. This feature has a fractional depth around -0.002; it has good signal-to-noise in terms of scatter between spectral channels, but is weak compared to several instrumental features in the band. Thus, it would be difficult to robustly identify any single molecular lines of similar depth originating from Venus' atmosphere, even at this telescope designed for minimal reflections. Types of problems are discussed next, with possible mitigations following in Section 3. 

\subsection{Instrumental issues}

\subsubsection{Sparse observations}

Spectroscopy of planets is often undertaken over a protracted period, either to build up signal-to-noise or for monitoring of temporal effects. When data are acquired in multiple sessions, instrumental ripples can shift in frequency and intensity. Figure 2 illustrates the discovery dataset of the absorption attributed to phosphine gas in Venus' atmosphere, acquired over a week in June 2017 (Greaves et al. 2021). The individual spectra are stacked vertically by time in the upper panel, with prominent ripples highlighted by bright/dark pixels. These pixels appear to travel across the frame, indicating artefacts of different phases over time. The lower panel shows the Fourier components corresponding to these ripples. Even in this small dataset (0.25 GHz band), there are four persistent ripple periods (hot pixels on the white horizontal axis). There are also some ill-defined temporal changes (hot pixels off this axis or in columns), in addition to the expected day-to-day periodicity (signals decline during each morning of observation, as the telescope's efficiency drops as it warms up). Well-filled temporal sequences of observations are often hard to obtain, making it challenging to model such instrumental behaviours in sufficient detail.

\begin{figure*}
	\includegraphics[width=2.1\columnwidth]{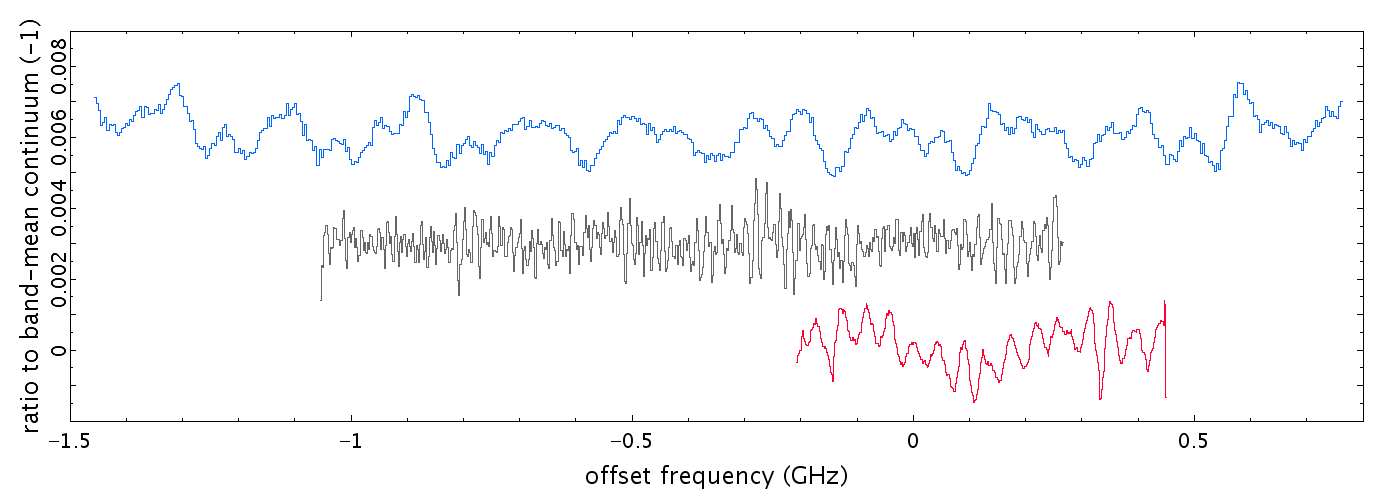}
	\includegraphics[width=2.1\columnwidth]{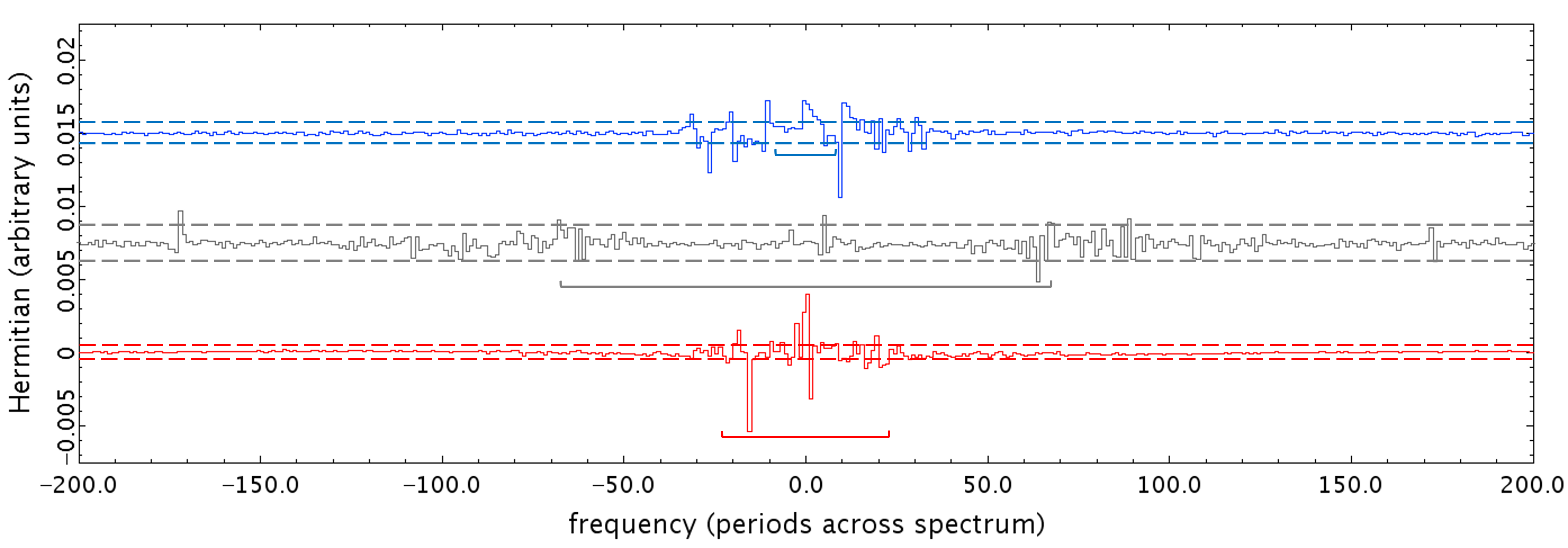}
    \caption{Upper panel: time-averaged spectra of Venus, offset vertically for clarity. From bottom to top: GBT, JCMT and SOFIA (scaled x0.1), using  their respective instruments KFPA, Namakanui/'\={U}'\={u} and GREAT/4G2. Offset frequencies are with respect to 24.0, 267.0 and 1067.0 GHz; these bands cover respectively the (J,K) inversion transitions of NH$_3$ and the J=1-0 and J=4-3 transitions of PH$_3$. Spectra were converted to fractional signals as described in the text. Data were processed from individual calibrated spectra for GBT and JCMT and a single Level 4 data-product was downloaded for SOFIA. Lower panel: Fourier components present in the spectra, in the same order as the upper panel and divided by 10 for SOFIA. Horizontal dashed lines show the $\pm$10$\sigma$ levels for each telescope. The bars with up-ticks at the ends below each Fourier spectrum indicate the ranges of low-frequency ripple periods that were suppressed by average filtering before the Fourier characterisation. Suppressed periods are calculated from the number of channels in the data (upper panel) divided by the number of channels over which the average filter was run. The small-period features for GBT occur because the processing did not completely remove a wide bowl-shaped feature (bottom spectrum in upper panel); this feature has $\ll$1\% of the original amplitude.}
    \label{fig:2}
\end{figure*}

\begin{figure*}
	\includegraphics[width=2\columnwidth]{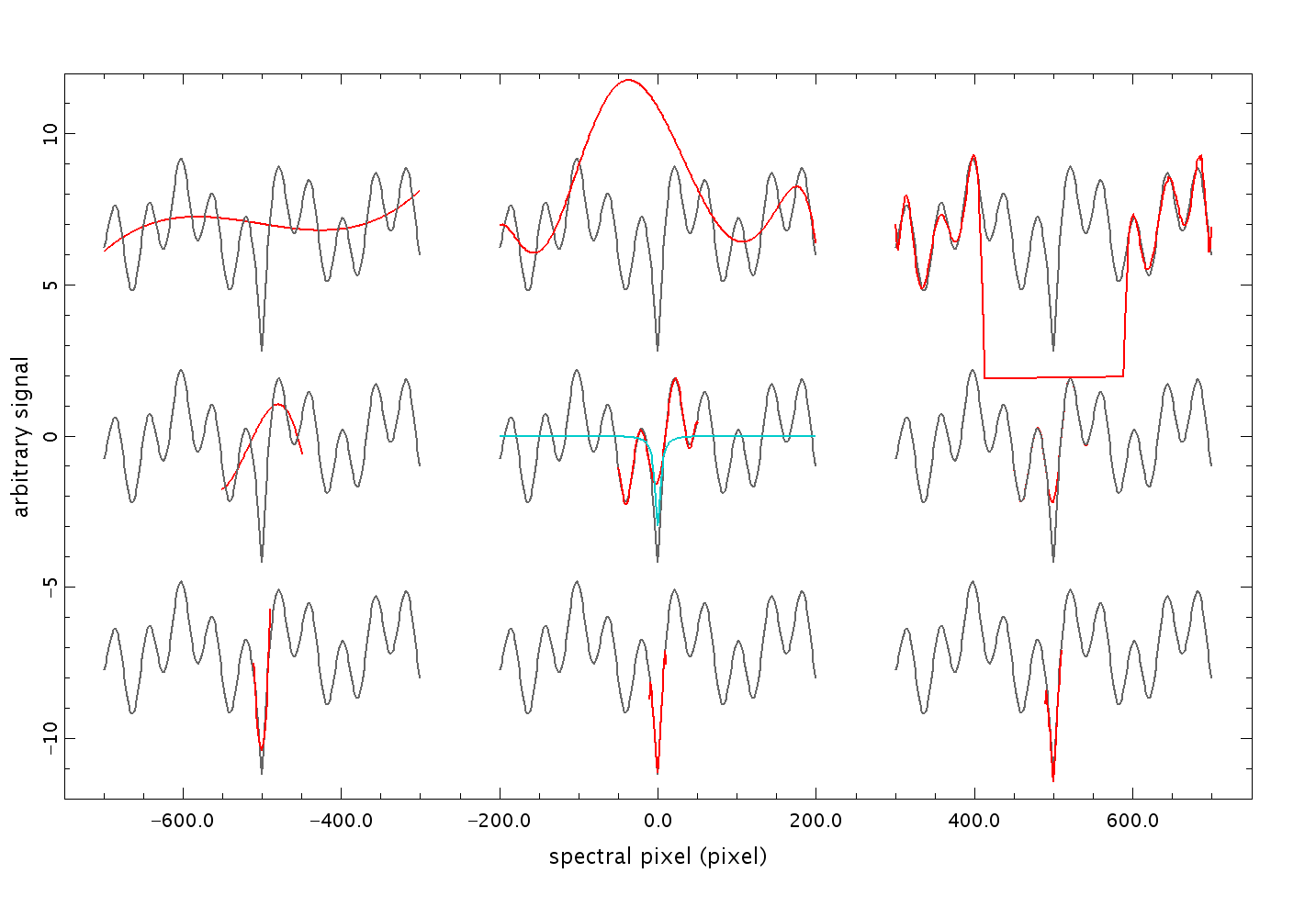}
    \caption{Results of polynomial fitting to simulated data. The ripple pattern (grey curves) is the sum of two pure sine-waves, to which is added a Lorentizian line model with a depth of 3 units (centre blue curve). The red curves show the polynomials in the regions where they are fitted; they have respective central windows (bridging regions) that cover 2, 16 and 200 pixels in the rows from bottom to top. The polynomial orders applied are 3, 8 and 14, in columns from left to right.}
    \label{fig:2}
\end{figure*}

\subsubsection{Ripples in the spatial domain}

When spatially-resolved observations are made of planets (usually with interferometers), the datacubes can show ripples in both spatial and spectral dimensions. Greaves et al. (2022) illustrated this in the case of the search for phosphine at Venus with ALMA in March 2019. Severe asymmetries were present in image-slices at each velocity, and there are smaller-scale artefacts that required high-order image-plane fitting to suppress. To illustrate the magnitude of ripple problems for very extended targets, Figure 3 shows an example spectrum from an ALMA observation of Jupiter. There is a broad high-amplitude ripple across the 1.6 GHz band, as well as possible residual striping within the image plane. Future observatories will also face these issues -- for example, Butler et al. (2004) note the problem of high dynamic range that SKA will have in the spatial dimension (>1000 for Jupiter). 

\subsection{Complexity of ripple patterns}

The main focus here is the characterisation of persistent sinusoidal patterns that need to be removed to reveal weak planetary lines. The examples compared here (Figure 4) are spectra of Venus with large band-widths, which may assist in precise characterisation (as many wave-repeats are present). These GBT, SOFIA and JCMT spectra were acquired over 1, 3 and 6 days respectively (in Feb 2022, Nov 2021 and May 2023), and each dataset amounts to about 1 hour of observation. The individual spectra have ripple patterns that appear similar between observations, so unweighted spectral averages were made. Flat fractional spectra were then formed by dividing by the band-mean signal, subtracting an average-filter (over 10-50 channels) to remove gradients, and sigma-clipping (at 2.5-4$\sigma$) to remove residual features (including any planetary lines). The parameter choices were empirical, and designed to make spectra as flat and free of non-periodic features as possible. Such harsh processing would not be typical in a real line search; it is applied here to illustrate the complexity of surviving instrumental features. The Fourier transforms of these spectra are shown in the lower panel of Figure 4. These transforms were made on the co-added spectra (6-16 individual observations per telescope), so any time-dependent effects will add to the number of Fourier features. Here there is no characterisation of periodicity over time (which can exist: Figure 2); improvements with temporal sampling are discussed below in Section 3.3.

All the time-averaged spectra of Venus exhibit complicated ripple patterns. Even for the GBT, where the off-axis design minimises large-$D$ reflections, there are 5 peaks in Fourier space that are well above 10$\sigma$. The SOFIA spectrum shows a more complicated family of Fourier components, also at larger amplitude (scaled x0.1 in Figure 4) than for the other two telescopes. SOFIA shows the lowest-frequency ripple confidently identified, with only 10 periods across the band. The JCMT spectrum is particularly notable for families of Fourier peaks, suggesting similar but not identical ripples were present in the individual spectra. Very short-period ripples are also present, with as many as 170 wave-repeats occupying the band. The clarity of this signal shows that features with only $\sim$10 channels per period can be readily extracted. No ripples occupying fewer channels were identified for any of the telescopes (the Fourier spectra are cropped accordingly, i.e. only noise is found for x-axis values above the 200-period limits of the plots).

In the JCMT case, the narrow ripples appear very periodic to the eye, suggesting successive subtraction of sinusoids would clean the data. Manual fitting of amplitude-modulated sinewaves was attempted for some JCMT spectra, but left large residuals. In practice, the eye could only assess fits with up to 3 sinusoids present, and the FT indicates that there are more numerous components. In conclusion, none of these deep planetary datasets is amenable to an easy cleaning process that leaves only real planetary absorption. 

\section{Data cleaning}

In the presence of such complex artefacts, it is very challenging to clean out the ripple patterns and leave real planetary lines. In particular, if one real weak feature exists in the band, it can be characterised as part of the ripple pattern with very little change to the Fourier components. This was noted by Cordiner et al. (2022, 2023) and Greaves et al. (2023) in their analyses of the 1067 GHz Venus spectra from SOFIA, and divergent conclusions were reached from only subtly different processing. The need to protect pre-specified spectral regions also hinders serendipitous discovery of unexpected molecules with transitions elsewhere in the band. 

\subsection{Polynomial fitting}

A traditional method of artefact suppression around a pre-specified line frequency applies automatic fitting and subtraction of a polynomial baseline. Software routines for this task have been available for at least 50 years (e.g. Cram 1974). High-order polynomials can be used where a complicated ripple pattern is present, to closely match local trends in the data and "bridge" smoothly over a window where a line is expected. This approach has a long history among long-wavelength observers; see e.g. Gordon \& Gordon (1975) for HI 21 cm spectra flattened with an 8th-order polynomial. The use of high orders seems less familiar at shorter wavelengths, where spectra can be more crowded with lines. One example application of high-order polynomials in the optical/IR regime is to fit galaxy spectra, and some limitations on reliability are discussed by Baldwin et al. (2018). 

The drawbacks of polynomial fitting are in robustness and computability. The polynomial incorporates no knowledge of the underlying behaviour of the artefacts, e.g. that they are a sum of sine-waves. Hence, the bridged section resembles a spline fit (piecewise interpolation with low-order polynomials), linking the data on either side of the line position with a smooth curve. (It can thus be informative to compare the results from a high-order polynomial fit to those from locally fitted splines or low-order polynomials.) The removal of nearby "bumps" in the spectrum is cosmetic, and a lower-order fit over a smaller spectral section should give similar results (e.g. Greaves et al. 2021 for the Figure 2 dataset). The flattening of bumps around the line does not necessarily imply that the bridging within the line region was correct, so there is a risk of a false-positive outcome, especially if there is an instrumental ripple narrower than the bridging window. Fitting a polynomial also becomes computationally intractable and unstable as the order increases, and so orders offered in software are typically <15 (orders above 12 gave no improvements here). Such a limit prevents the extraction of wide low-amplitude planetary lines, if many ripple periods are contained within the line width. Fitting inside the line is also undesirable as real features could be removed; for example, a line may have both wide and narrow components if the absorbing molecule exists at a range of pressures.

\begin{figure}
	\includegraphics[width=\columnwidth]{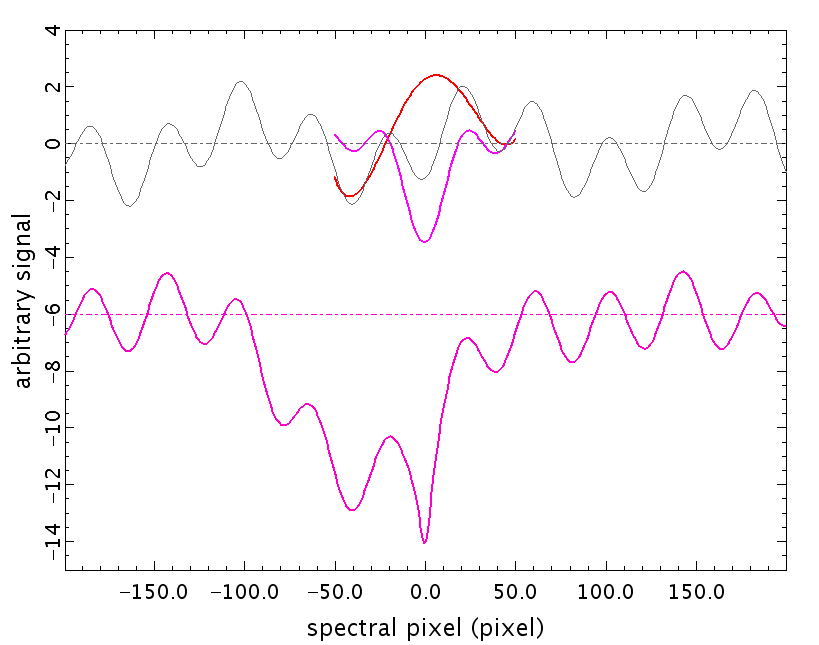}
    \caption{Top: Residual (pink curve) after subtraction of a polynomial fit (red curve) from the simulation as in Figure 5 but with no injected line (thin grey curve). The polynomial has order 4 and a bridging window of 30 km/s. Bottom: residual spectrum (offset vertically for clarity) after subtraction of the polynomial from the simulated data in the upper-middle case of Figure 5. }
    \label{fig:2}
\end{figure}

\subsubsection{Pitfalls}

Figure 5 illustrates some classic failures of polynomial fitting, resulting from poor parameter choices. Good choices set the polynomial order from the number of turning-points of the ripple pattern and place the bridging window closely around the target line. In Figure 5, wide windows (top row) result in inadequate information to model the ripple pattern and hence random outcomes, while narrow windows (bottom row) result in the line being mostly removed. The fitting is less dependent on the polynomial order (increasing left to right), but a low order leaves residuals while a high order can over-model the region around the line. In this simulation, the example at centre is most correct, with the difference of the data (grey curve) and fit (red curve) matching well to the injected model line (in blue). In real cases, where the expected line strength is often poorly known, exploring small variations in fitting parameters is important to avoid an overly-subjective "best result". 

The highest chance of a false positive comes from fits where a large vertical offset of the polynomial above the data forces a large negative feature after subtraction (lower spectrum in Figure 6). In this case, the error is recognisable because the residual "line" has an unphysical asymmetric profile. A more dangerous case occurs (upper spectrum in Figure 6) when there is no real planetary line, but a valley in the ripple pattern occurs near its frequency. Polynomial subtraction can then result in exaggerating the depth of the feature and shifting it towards the centre of the bridging window (target frequency), creating a more significant "fake line". In the noise-free simulation, the feature created does not have the profile of a true line (Figure 5). However, in real observations, the unphysical profile of the line-candidate may be hard to discern, especially in cases where real lines and ripples have similar widths. 

\subsubsection{Testing for fake lines}

The JCMT dataset is used to test whether fitting polynomials to small sections of the spectrum can result in frequent false positives. New instrumentation for the May 2023 observations provided a bandwidth of 1.8 GHz (compared to only 0.25 GHz in Figure 2), and 13 short test-observations were made. The HDO 2(2,0)-3(1,3) absorption line was the only feature expected to be visible in a co-added spectrum, after a total time on Venus of only $\approx$30 minutes. This feature was selected to "train" the polynomial baseline fitting procedure, to determine whether similar but fake lines can be generated. 

The Venusian HDO line at 266.161 GHz is weakly recovered with a 12th-order polynomial fitted over a 100 km/s interval with a bridging window of 10 km/s (Figure 7, middle-left panel). This window was chosen as similar to the period of the narrowest ripple present at JCMT; polynomials can not fit accurately for a wider window with instrumental structure within it (though wider HDO absorption may in fact be present). The high order allows good characterisation of the crowded short-period ripples. Higher orders negligibly reduced the noise in the fit-subtracted spectrum (by $<$10\%), while orders 6-11 showed larger residual structures (and were noisier by up to $\sim$25\%). The centre of the HDO line is offset by two spectral pixels (each of 0.55 km/s) from the expected position, which is attributable to low signal-to-noise. 

\begin{figure*}
	\includegraphics[width=2\columnwidth]{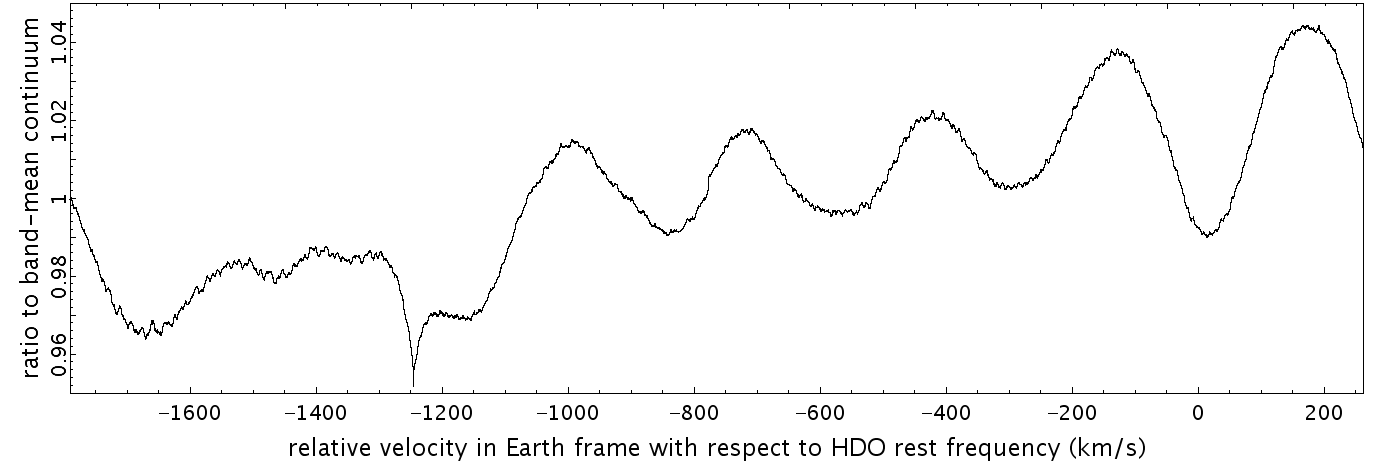}
	\includegraphics[width=\columnwidth]{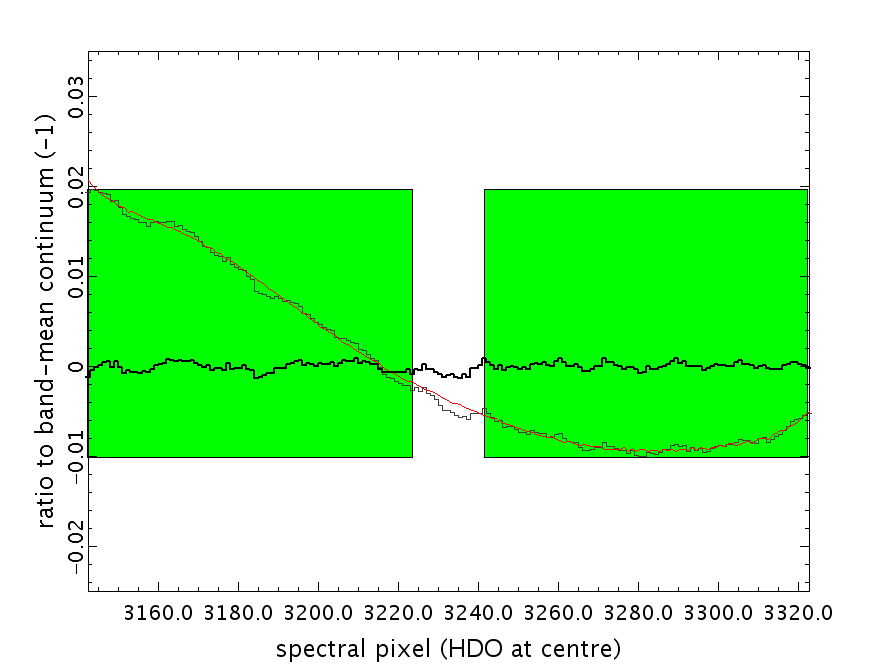}
	\includegraphics[width=\columnwidth]{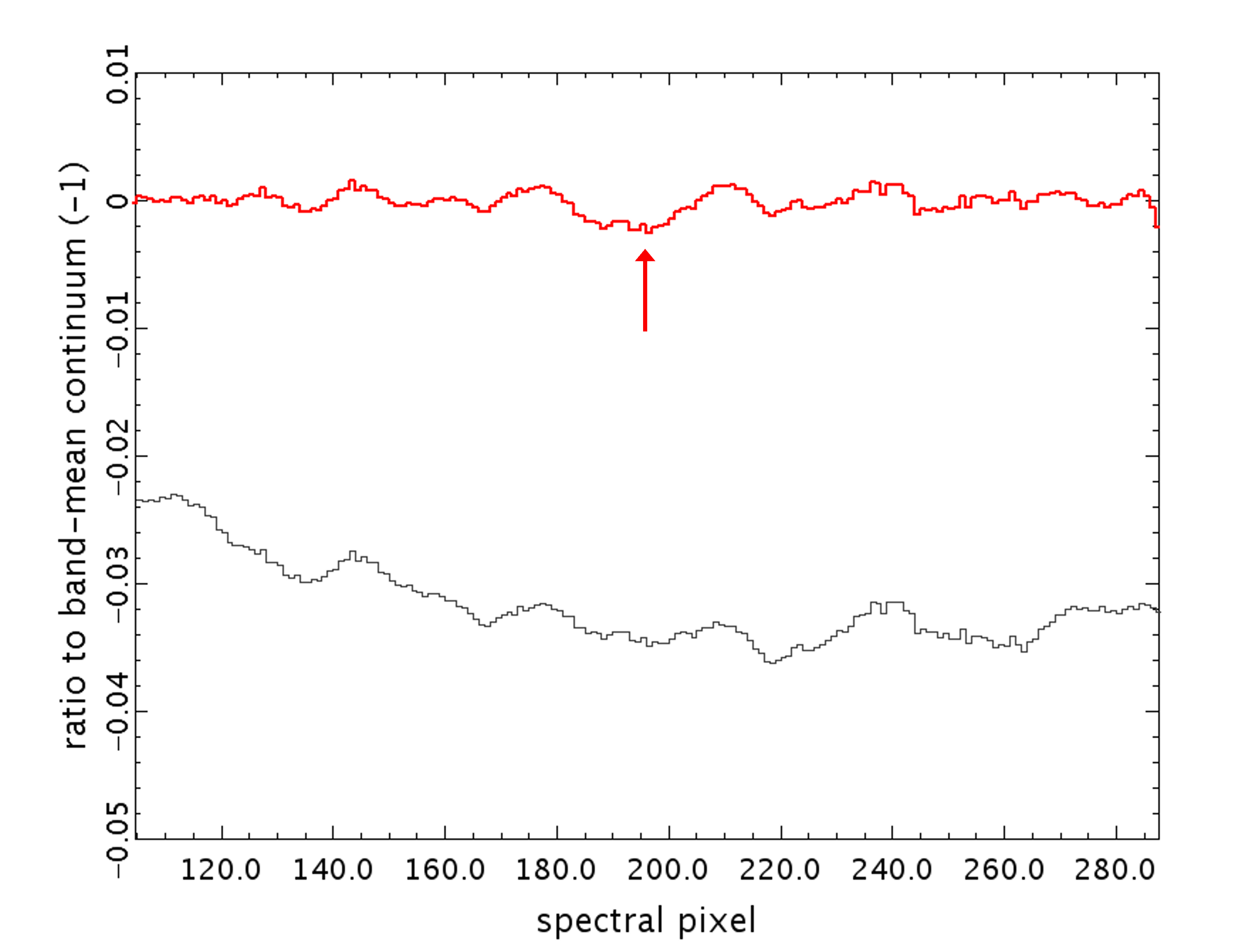}
	\includegraphics[width=\columnwidth]{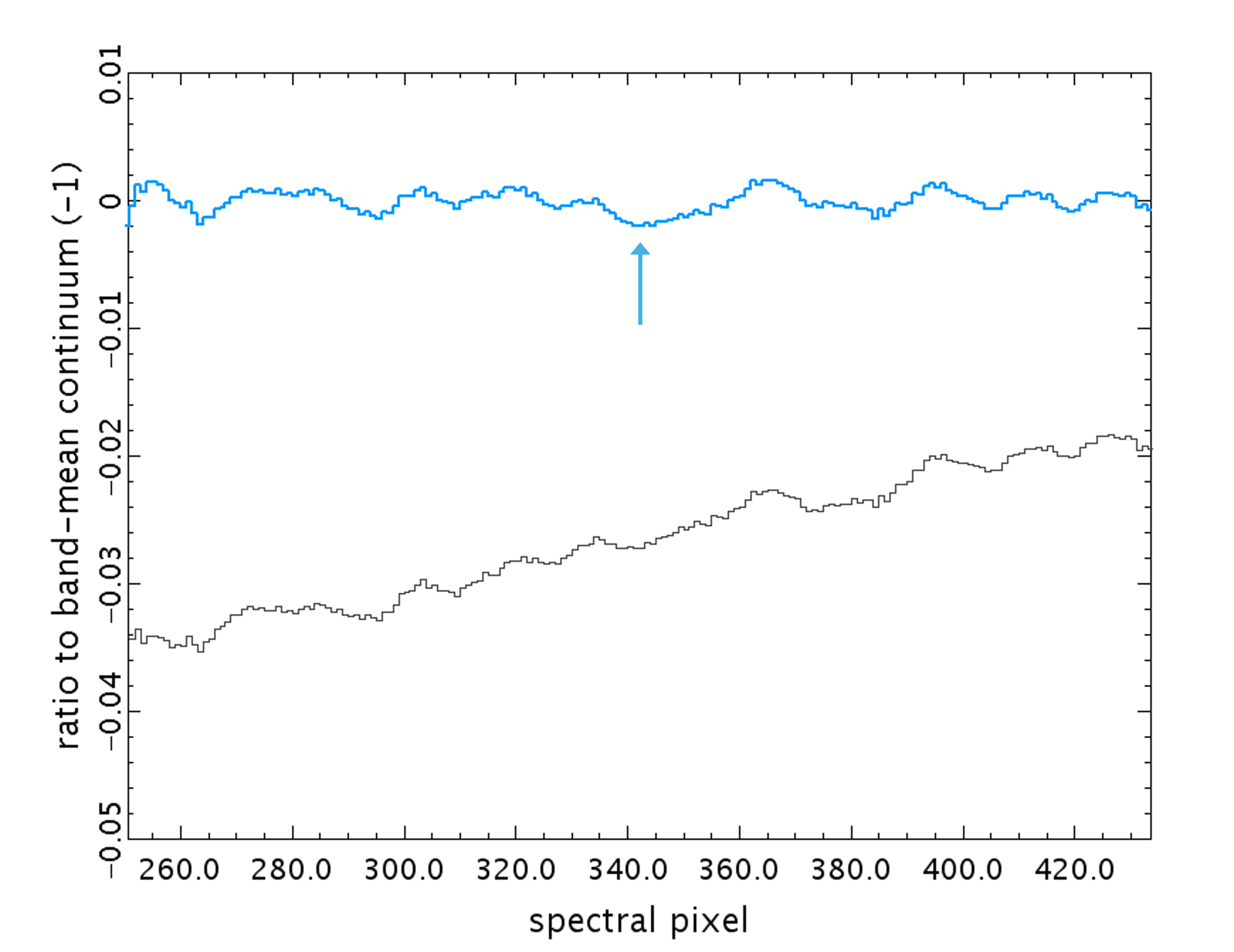}
	\includegraphics[width=\columnwidth]{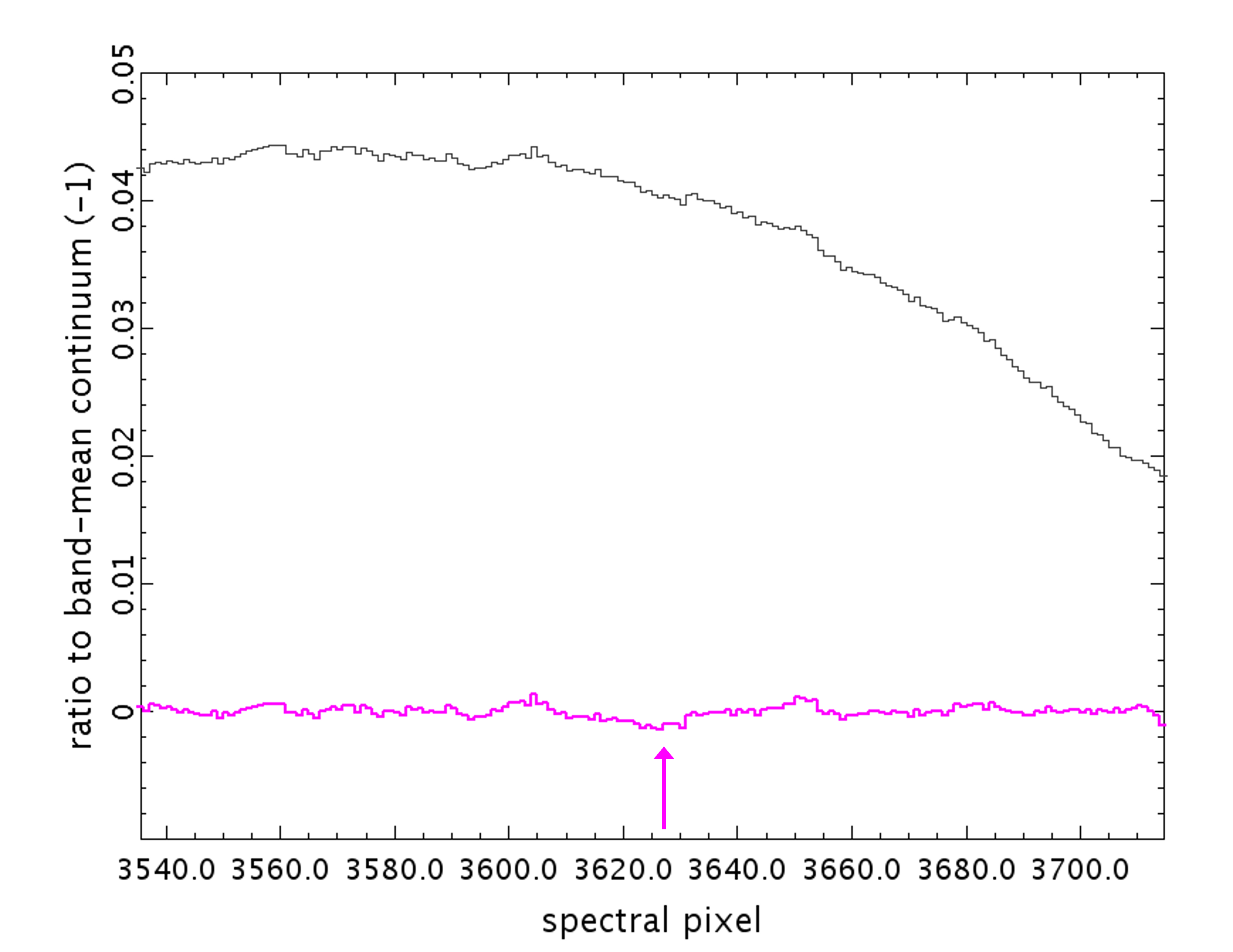}
    \caption{Results of the tests for production of fake lines by polynomial fitting, for the JCMT dataset of May 2023. Top: co-added spectrum on a velocity scale, with respect to HDO 2(2,0)-3(1,3) at a rest frequency of 266.161 GHz; Venus' velocity was $\approx$-13.65 km/s with respect to the telescope. A terrestrial O$_3$-residual leads to the feature near -1250 km/s. Middle-left: fitting process for the candidate HDO feature, with the 12th-order polynomial in red, bridging over the gap between the two green boxes; the original and fit-subtracted spectra are shown with thin and thick black histograms. Remaining 3 panels: search for fake lines using the same fitting procedure, as described in the text, with candidate features (red, blue and pink histograms) appearing centrally on the x-axis. The full range of the spectrum covered 3731 pixels, and these three most-viable fakes occur near the band ends. The central velocities of the 3 features are -1682, -1602 and +204 km/s, with higher pixel number corresponding to increasing velocity.}
    \label{fig:1}
\end{figure*}

To test for fake lines similarly close to pre-specified positions, the 100 km/s fitting-interval was stepped across the entire passband, by 2 spectral channels (1.1 kms) at a time. Candidate fake lines were retained for further inspection if the polynomial fit produced a line-minimum within 1 km/s of the window-centre. The total number of possible tests under this procedure was $\approx$1750, in a band $\approx$1940 km/s wide. Some adjacent steps produced very similar outcomes (due to the small velocity shift) and these duplicates were not retained. 

\begin{figure*}
	\includegraphics[width=2.1\columnwidth]{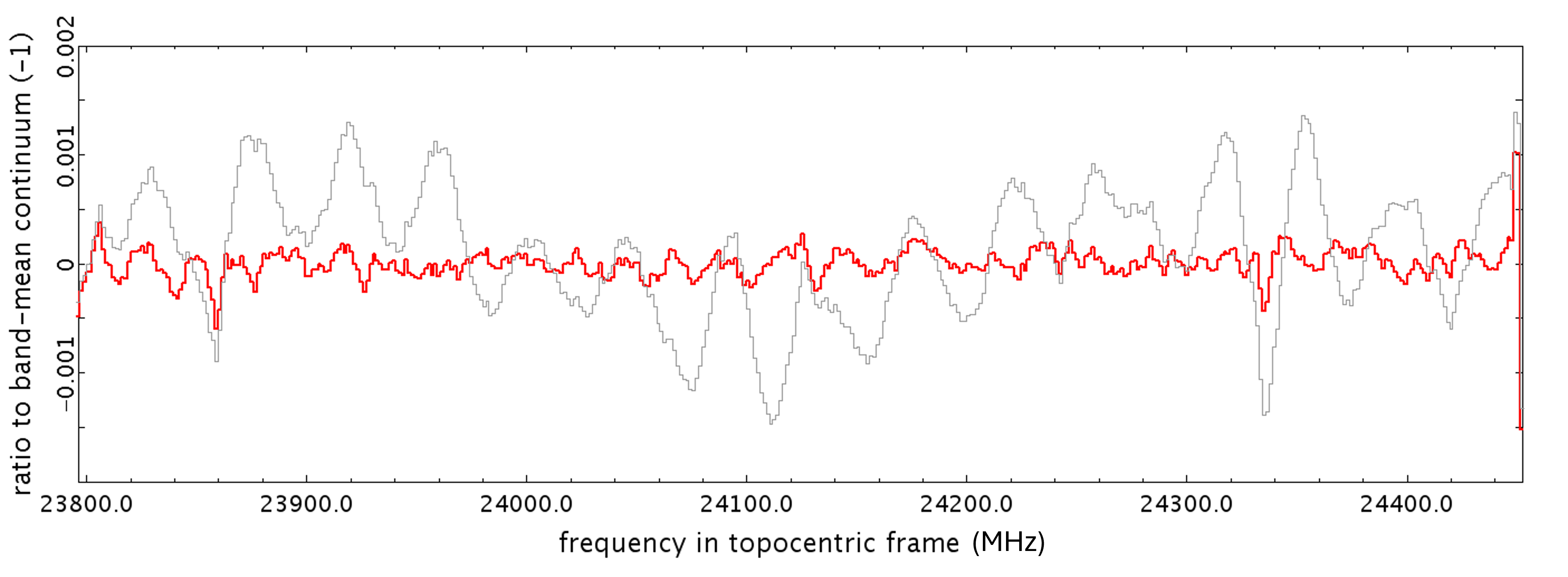}
    \caption{GBT spectrum as in Figure 4 (grey histogram, overplotted after cleaning the 10-sigma Fourier components as described in the text (red spectrum). The absorption line-minimum in the channel at 23859.1 MHz is identified with terrestrial ozone absorption at a rest frequency of 23859.6 MHz (channel spacing in this plot is 1.465 MHz).  }
    \label{fig:1}
\end{figure*}

A dozen fake-line candidates were generated by this process, for an initial false-alarm rate of $\sim$0.7\%. On inspection, most of these were "clones" of the original shape of the spectral baseline, i.e. a 1st-order baseline fit would be sufficient to recover them. Such cases would always be suspicious and need independent confirmation even if they did match a known line frequency. More interestingly, the candidate set included three features that appeared more distinctly after the high-order polynomial fitting (Figure 7). However, further tests showed that these three fake-line candidates would not pass as robust real features. Their suspicious characteristics include asymmetric or rectangular line-shapes; having a realistic line profile only at one polynomial order; or becoming significantly deeper as the polynomial order increases. All three spectral sections were also near the ends of the 1.8 GHz band, where acquired signal drops off leading to worse robustness. These three results are example false positives; a lower noise level would be needed for a false-negatives test, i.e. for any additional real lines to be significant above the noise. The fake-line test could also be re-run after fitting different orders of polynomial, but similar orders tend to generate similar false-positive features.

In this particular test, the three fake-line candidates occupy only three 10-km/s sections of the $\approx$1940 km/s band, so the false-alarm probability is low. Given that none of these candidates survived further scrutiny, the false-alarm probability is formally 0 of $\approx$1750 tests. An achievable goal for most GHz-THz planetary observations with similar bandwidths\footnote{Pressure broadening of lines, described in units of e.g. MHz/bar, tends to produce lines of similar width in MHz for diverse wavelength regimes.} would be to test for zero plausible fake lines in >1000 tests.

\subsection{Fourier-based ripple suppression}

If the artefacts present in observed spectra are all pure sinusoids, then it should be straightforward to completely characterise and subtract them, producing results limited only by noise. A frequency-space approach has been implemented for example at the MOPRA 22m (Walsh 2009). In such a Fourier space, even a low-amplitude ripple should show a clear signal, as many repeats are present within a wide band. This is confirmed by the clear detection of waves of fractional-amplitude <0.001 that repeat 170 times within the JCMT band (Figure 4). 

The simplest ripple pattern of the three telescopes tested is that of the GBT bottom red spectra in Figure 4). A model for the GBT spectral baseline was constructed by isolating the Fourier components that are 10-sigma above the noise, measuring this standard deviation towards the ends of the Hermitian spectrum. Even at this very conservative cut (well above the noise), there are four distinct sine-waves present, plus some power spread among low Fourier-space frequencies. These >10-sigma components were then all inverse-Fourier transformed to create a model spectral baseline, which was subtracted from the data. The full band was used for the Fourier analysis, i.e. no spectral regions were "protected".

Figure 8 shows the result compared to the original spectrum. The strongest feature is now the terrestrial ozone line at 23.8596 GHz, as seen against the Venus continuum. This line is within <1 spectral channel of the expected position and the line width has not been altered by the processing. A real feature has thus been successfully identified, although some other residuals are of almost equal magnitude. The standard deviation of the whole spectrum has been reduced from 0.0006 to 0.0001, a factor of six. This is a modest gain in dynamic range, but could be  scientifically important if this were a planetary rather than terrestrial feature. The fractional depth of this O$_3$ residual is near the limit achieved in previous planetary spectra (e.g. Matthews et al. 2002). 

\begin{figure}
	\includegraphics[width=1.05\columnwidth]{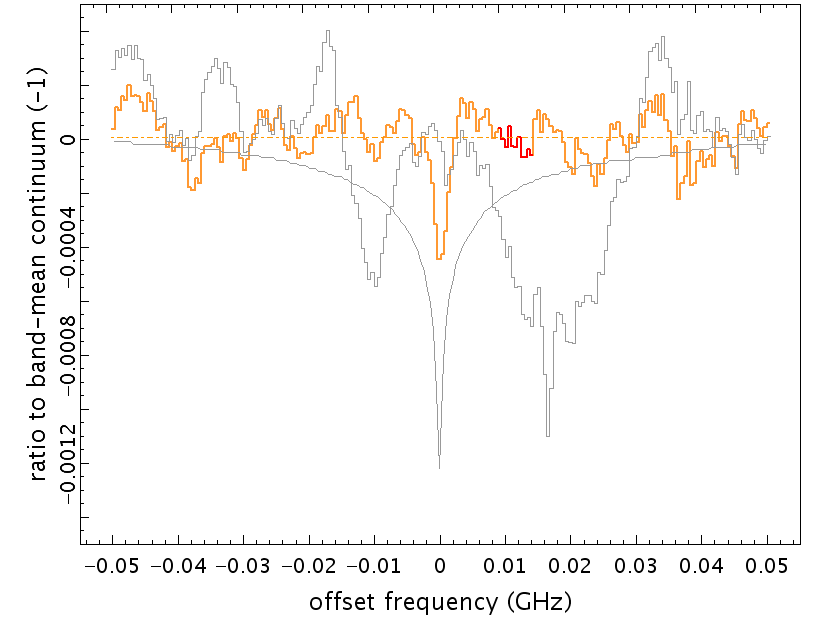}
    \caption{Simulation of recovery of sulphuric acid vapour lines, as described in the text. The grey histogram represents the co-add of 10 spectral sections around the rest frequencies of 10 H$_2$SO$_4$ line components, in the JCMT dataset of May 2023 (Figure 7, but now also including the data acquired in the opposite sideband). The grey curve is a model for 10 ppb of H$_2$SO$_4$ vapour in Venus' atmosphere, which was injected into the real data. The orange histogram shows the partial recovery of the injected feature after Fourier-cleaning. The red section of this spectrum indicates where a real H$_2$SO$_4$ vapour signal from Venus' atmosphere would lie, for the frequency offset from rest-frequency at the time of observation.}
    \label{fig:1}
\end{figure}

\begin{figure*}
	\includegraphics[width=1.55\columnwidth]{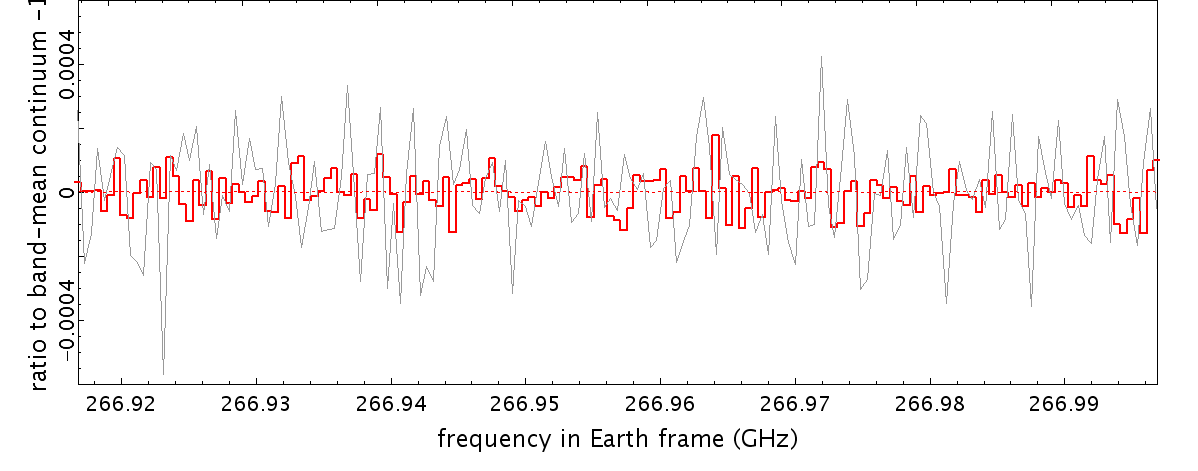}
        \includegraphics[width=0.51\columnwidth]{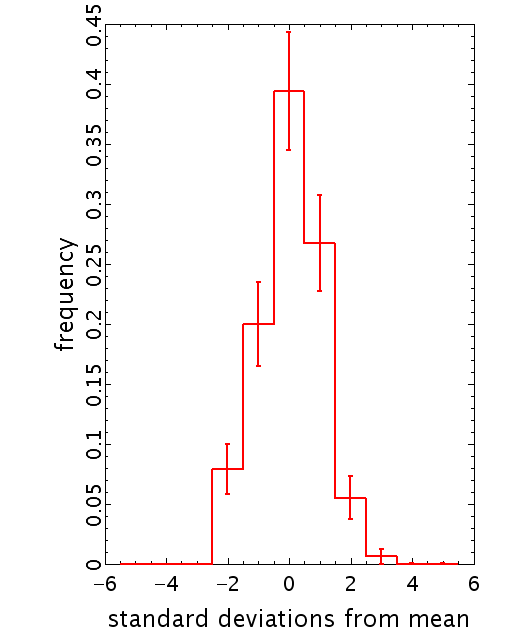}
    \caption{Left: comparison of Fourier-filtered spectra with the input data processed separately versus as an average, for the May 2023 JCMT run. The data are for one spectral window and one polarization, and using only the first observation acquired on each of the 6 evenings. The grey line-plot shows the result of Fourier-cleaning the co-add of the 6 spectra. The red histogram shows the co-add of the cleaned results for each of the 6 spectra. A 2.2$\sigma$ feature is present (red spectrum) near where Venusian PH$_3$ J=1-0 absorption line would be expected, at 266.957 GHz, Doppler-shifted from the rest frequency of 266.945 GHz. Right: statistics of the red spectrum from the left panel; uncertainties are from Poisson statistics. The weak feature near the PH$_3$ frequency was included (4 channels, 2.5\% of the data). }
    \label{fig:1}
\end{figure*}

\subsubsection{Fidelity in line recovery}

In Figure 8, the single narrow O$_3$ line would not be expected to be filtered out in Fourier processing, as it has a narrower width than the instrumental ripples. Figure 7 illustrated a more challenging case, where the single HDO line has comparable width to the ripples seen at the JCMT. The effects of different techniques for ripple removal on this line will be discussed in Tang et al. (in prep.), using an extended dataset from the JCMT-Venus Legacy project\footnote{https://www.eaobservatory.org/jcmt/science/large-programs/jcmt-venus-monitoring-phosphine-and-other-molecules-in-venuss-atmosphere/}. Here, a simple test is made of fidelity under Fourier processing, by recovering synthetic lines injected into the JCMT test data of May 2023. 

This dataset has two sidebands (centred at 255.045 and 266.845 GHz), and a total frequency span of $\approx$3.5 GHz. A use-case for this wide dataset is the search for sulphuric acid vapour in Venus' upper atmosphere (mesosphere), as H$_2$SO$_4$ has 10 similar absorption components across the two sidebands, at gradually increasing separations (from 380 MHz up to 435 MHz between adjacent lines). These components thus form a non-linear comb of narrow features and so the prior expectation is that they should not be removed in Fourier processing. 

For the injection test, a model spectrum for 10 ppb of H$_2$SO$_4$ in Venus' atmosphere was created using the online Planetary Spectrum Generator (PSG\footnote{https://psg.gsfc.nasa.gov/}) tool. For simplicity, one representative model was generated for the 10 similar features (using the 255.786 GHz line), and this was cloned with the clones injected into the real spectrum at the 10 rest frequencies. This injection was run before the average-filter step, for which the window was reset to 64 channels (tests showed the line-minimum depth was then well retained). The dataset was Fourier-processed as above, with a 10-sigma cut in Hermitian space, and no protection of line regions. Ten sections around the H$_2$SO$_4$ rest frequencies were then cut out, converted to offset-frequency, and co-added with equal weights. 

Figure 9 demonstrates that the net spectrum has been significantly improved in dynamic range by the ripple suppression (comparing the grey and orange histograms). The injected feature is recovered with high confidence, and fractional noise-residuals are only $\approx$0.0002 at peak. Co-adding the 10 non-periodic spectral sections has likely helped to suppress instrumental residuals. However, the injected net feature has been reduced in depth by around two-thirds after the cleaning process. Thus the majority of the test-signal has been lost, in spite of the comb pattern of injected lines not being periodic. A more severe 3-sigma cut in Fourier space only marginally improved the dynamic range but led to severe signal loss, of >80\% of the original injected line-depth. 

Figure 9 shows only noise where real Venusian H$_2$SO$_4$ absorption would appear (red section of the spectrum). A 3-sigma limit in this section corresponds to approximately 3 ppb of sulphuric acid vapour, assuming that the proportion lost in Fourier-cleaning is similar to that of the injected lines (10 ppb model). This is the same abundance limit as achieved by Sandor et al. (2012), but in their case from a multi-year campaign at JCMT with narrow-band instrumentation. Automated processing of new wider-band data is therefore a potentially powerful technique, given that the on-source time at JCMT in May 2023 was only about an hour. 

\begin{figure*}
	\includegraphics[width=1.42\columnwidth]{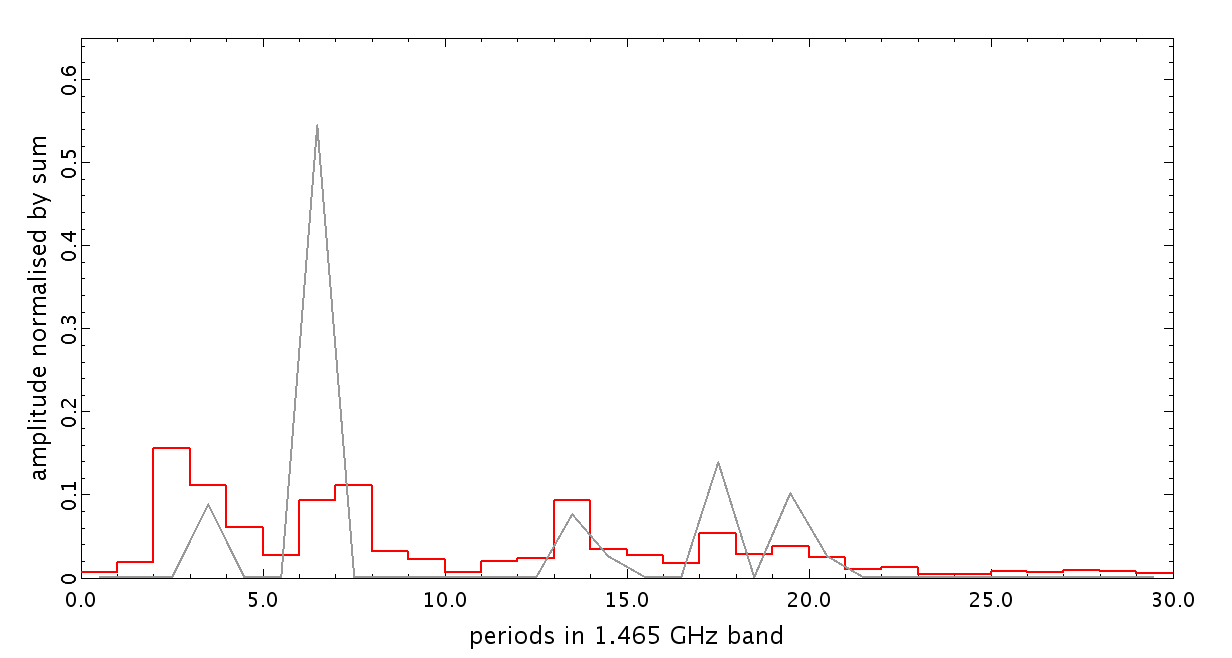}	
        \includegraphics[width=0.57\columnwidth]{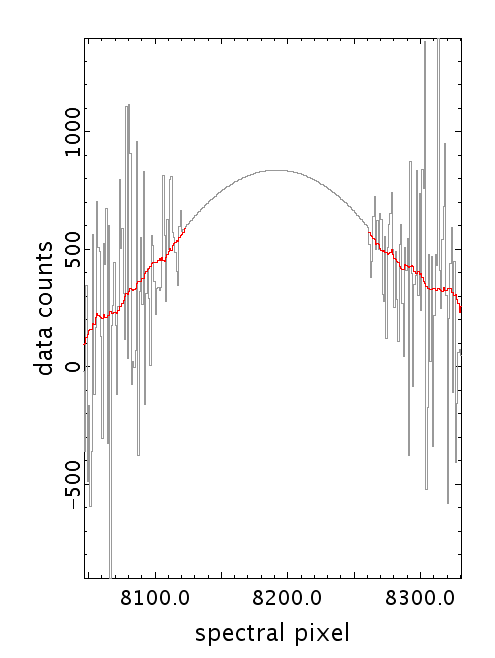}
    \caption{Left: Comparison of periodic artefacts found in the 1 THz SOFIA spectrum of Venus (Figure 4), using Fourier transform versus Lomb-Scargle periodogram analysis. The delta-functions (non-zero points in the grey line plot) correspond to the 7 Lomb-Scargle components fitted and subtracted in Figure S2 of Cordiner et al. (2022). Amplitudes and frequencies in radians/GHz were read off the figure, and are re-plotted here as normalised amplitude against periods across the band analysed by Greaves et al. (2023). The red histogram illustrates the normalised amplitudes found for Fourier components in Greaves et al. (2023); pre-processing of the data was different to that of Cordiner et al. (2022). Right: illustration of the interpolation needed in Fourier processing, to bridge across a line (here one PH$_3$ J=4-3 component in a single spectrum). The grey histogram shows the data with the quadratically-interpolated section and the red histogram shows the data trend (average-filter run over 50 spectral channels).    }
    \label{fig:1}
\end{figure*}

\subsection{Temporal sampling}

For sparsely sampled data, instrumental effects can change (Figure 2). Figure 10 illustrates how a Fourier cleaning process can be optimised by running it for separate observations and then averaging the cleaned spectra. To isolate effects that changed from day to day within the test observations, only the first spectrum acquired each evening in May 2023 was used (a few minutes of Venus observation in each case). The Fourier cleaning adopted a 10-sigma cut as before. The 6 spectra were cleaned and then co-added (red histogram), for comparison to the result (in grey) of co-adding the six spectra and then cleaning the average. 

Cleaning the individual spectra significantly reduces the "spikiness" of the output (Figure 10). This suggests that periodic ripples in the data are better removed by treating them over shorter timescales. If the data are all averaged, time-varying ripples are combined into one very complex pattern, and so the conservative 10$\sigma$ cut in Fourier space has likely not found all of the features. The dynamic range improvement is such that a candidate line is now weakly visible (centre of Figure 10), even though the on-source time was only around 0.5 hours and this line region was not protected in the processing. This is encouraging for fully-automated approaches, although this weak feature is only at 2.2$\sigma$ when summed over 4 channels, or 1.4\% probability by chance if the noise is normally distributed. At this confidence level, similar artefacts are likely, and in fact a 2.9$\sigma$ feature appears at the far-right of Figure 10. 

At least for this telescope, it appears that instrumental ripple patterns are not fully repeatable over time, but are predominantly periodic in single spectra. For narrow lines, an automated time-separated Fourier-cleaning approach can then suppress most artefacts in the spectrum, to leave noise that is close to normally distributed (right panel of Figure 10). For GBT and SOFIA, there was fewer spectra over more limited time-periods, and no strong temporal changes were identified. 

\begin{figure*}
\includegraphics[width=2.1\columnwidth]{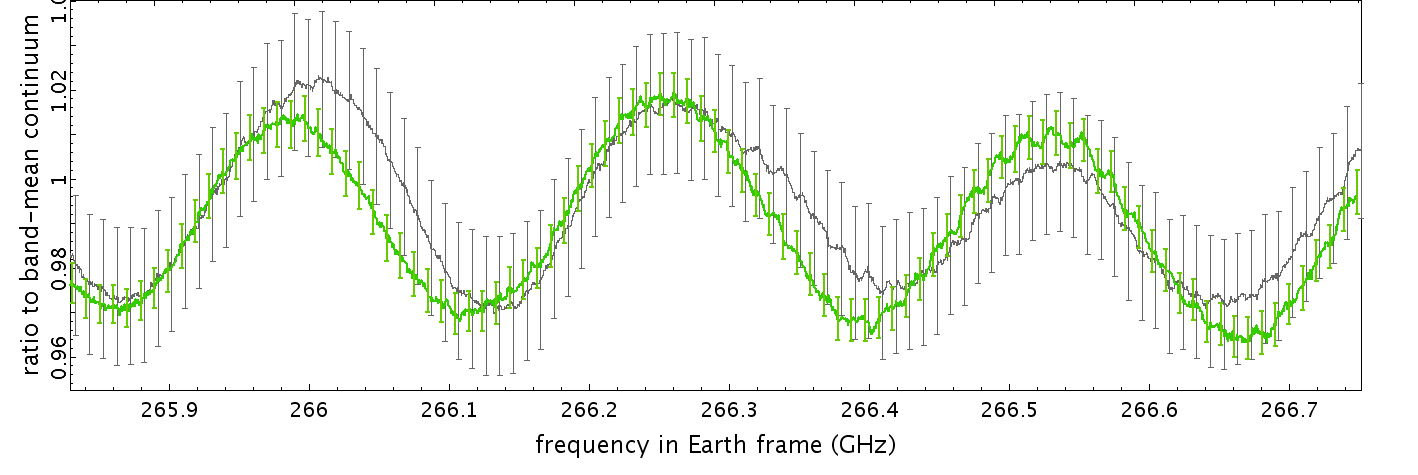}
    \caption{Test of mitigation in hardware. The grey spectrum of Venus was observed on 1 July 2023, and is the unweighted average of 5 similar observations. No average filter has been applied.  Standard errors are plotted (at every 20th spectral channel, for clarity), and were generated from the dispersions of the input data. The green spectrum is the equivalent for 5 observations on 4 July 2023, after adding some eccosorb (microwave-absorbent material) around the receiver.  }
    \label{fig:1}
\end{figure*}

\section{Discussion}

\subsection{Comparison of methods}

There has been little prior benchmarking of different ripple removal techniques. Figure 11 shows a test of the Fourier approach compared to that of Cordiner et al. (2022) in their analysis of the 1 THz SOFIA spectrum (Figure 4). These authors used a Lomb-Scargle periodogram technique, where sine-waves are successively cleaned out; they expand on some advantages of this method in Cordiner et al. (2023). An especially useful feature is that the LS-periodogram can work with missing data, i.e. pre-specified line positions can be naturally protected. In contrast, Fourier processing needs to bridge over line regions using suitable local substitutes for sinusoidal sections. For this dataset, up to quadratic orders were adequate (right panel of Figure 11).  The same spectral regions were protected by both Cordiner et al. (2022) and Greaves et al. (2023). 

Figure 11 illustrates that the Fourier analysis and the LS-periodogram are identifying ripples with similar periods. The broader ripples (fewer periods across the band) appear somewhat less clean in the Fourier analysis, i.e. power is spread among more x-axis bins than for the periodogram. The Fourier peaks are also smaller in amplitude than the largest LS-periodogram component. This difference likely arises because the LS-cleaning is iterative, and the largest component may also clean out some signal at multiples of its ripple-period. 

The two processing approaches therefore appear quite comparable, and they were similarly effective in improving dynamic range (Greaves et al. 2023). However, there is still debate over the fidelity of candidate features of phosphine (Cordiner et al. 2023). 

\begin{figure*}
\includegraphics[width=2.2\columnwidth]{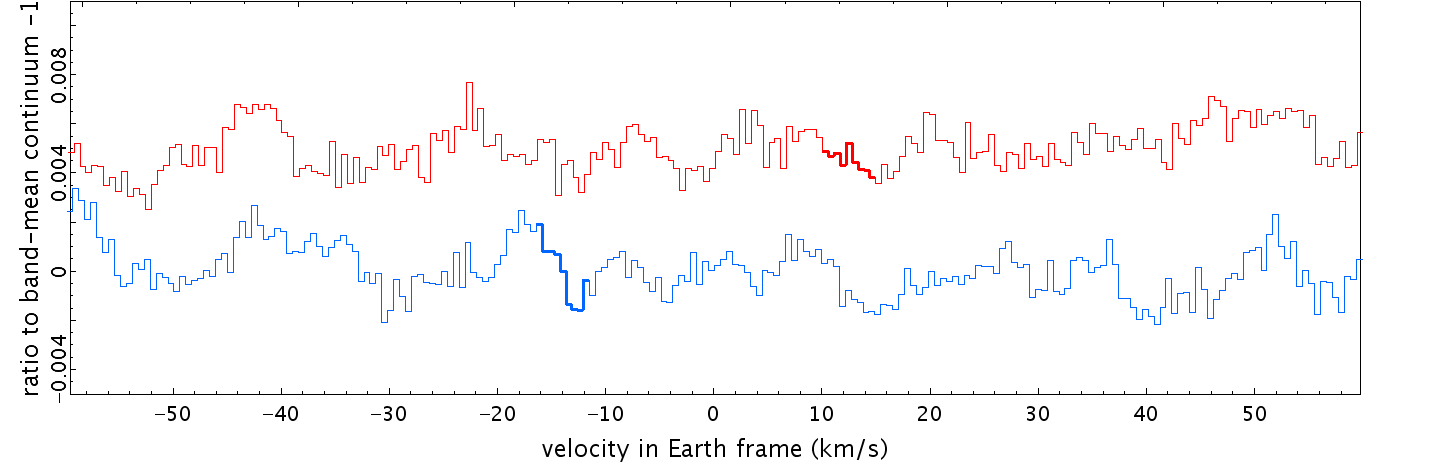}
    \caption{Illustration of a search for a planetary feature using relative motion as a robustness test. Venus was observed at the JCMT on 26 May 2023 (blue histogram) and 24 Sep 2023 (red histogram, offset vertically for clarity), when the planet was respectively approaching and receding from the Earth. The reference frequency for the spectral axis is 266.483 GHz, corresponding to one of the H$_2$SO$_4$ line-component rest frequencies. The velocities to which a Venus signal would be Doppler-shifted on these dates are indicated by the segments shown in thicker lines. The spectra each consist of the co-add of 3 observations, with average-filters subtracted.   }
    \label{fig:1}
\end{figure*}

\subsection{Limitations}

Simple data-cleaning interventions have been shown here to have positive outcomes, with dynamic range improved by an order of magnitude (e.g. detectable line-fractions now approach 10$^{-4}$; Figures 9 and 10). Fourier-cleaning can be done in a single step, and can handle  ripples that are not pure sine-waves (occupying a few bins in Fourier space, Figure 11). Lomb-Scargle periodograms function similarly, but results can be examined at iterative steps, and protecting line regions is more straightforward. In either approach, the  remaining challenge to dynamic range improvement would be the presence of artefacts that are not periodic -- for example, single "lumps" in the spectrum that could mimic real lines. Figure 7 showed some strategies for testing the frequency of such false positives. Though rare, even single instances can affect conclusions about whether a line detection is robust. For example, in Figure 8 there is an absorption-like feature near 24.34 GHz, which is almost as significant as the terrestrial O$_3$ line at 23.86 GHz. 

\subsection{Mitigation}

Mitigation in hardware is challenging because of the complex nature of reflecting surfaces around an instrument. Figure 12 illustrates an attempt at minimising reflections made at the JCMT. Microwave-absorbent material (eccosorb) was used to clad some of the surfaces around the receiver, aiming to suppress some of the short-distance (large-period) ripples. Ripple periods of around 300 MHz suggest reflecting surfaces $\approx$50 cm away, but these could not be exactly identified (the receiver is in a cabin with numerous struts and panels). Mitigation methods used at other telescopes are referenced in Section 1.2.

Figure 12 suggests that the amplitude of the broad ripples was not in fact significantly suppressed in these tests (green versus grey spectrum). However, the ripples were more reproducible between spectra than before the cladding was attached -- the green spectrum has smaller internal dispersion. The reduction in scatter is a factor of a few, but this is not well-quantified as it is for 5 observations only in each case. The ripple pattern in the green spectrum in Figure 12 also appears more symmetric -- i.e. closer to a pure sine-wave -- than in the grey spectrum. More tests would be useful to see if these conclusions hold up over longer periods of observation. 

Narrow ripples are more problematic for many studies. One approach to robust detection would be to re-observe when the target has shifted significantly in velocity with respect to the telescope. Figure 13 illustrates this approach, using data taken 4 months apart, around the main runs of the JCMT-Venus Legacy program in 2023. The observing dates were near the maxima in Venus-Earth relative velocity, with Venus first approaching and then receding. The highlighted portions of the two spectra illustrate where the same H$_2$SO$_4$ line would be located for the two dates. The Doppler-shift of 26 km/s is sufficient for a Venusian line to cross three periods of the narrowest ($\approx$8 MHz) ripple in the data. 

This test offers some robustness, as the real planetary line occurs in a different place in the ripple pattern. However, the ripples were not fully repeatable over such a long time period (Figure 13). Tang et al. (in prep.) will discuss JCMT results with better time-filled data, when Venus' velocity changed by around 3 km/s over 3 weeks. 

More generally, implementing the Doppler-shift approach needs careful scheduling for other solar system targets -- for example, Mars' velocity changes more slowly with respect to Earth, while Titan's changes quickly in its orbit around Saturn. Repeat scheduling may not always be practical, for example if the target moves outside the nightly observing hours, or an instrument configuration is changed. 

\section{Conclusions}

The analyses of spectra of Venus acquired with GBT, JCMT and SOFIA demonstrate that periodic ripples in spectra form very complicated patterns. Even within wide bands, the overall pattern formed by a few uncorrelated sinusoids may never repeat, and so the direct comparison of a line-region and counterpart line-free region can be impossible. However, wide bands allow for strong testing of polynomial-fitting methods, e.g. that no line-free regions produce fake lines in >1000 tests. This requires sufficient signal-to-noise for the centre of the candidate line to be characterised to better than one-thousandth of the bandwidth. Over 1000 tests stepped by this interval can then be run; spacing the tests more closely than this uncertainty yields no additional information.

Overall, we conclude that, as polynomial fitting does not have predictive power to say what the model spectral baseline should be across the line region, characterising and subtracting the underlying periodic ripples can be more powerful. It is useful to inject synthetic lines into the data to check what signals remain after data-cleaning, as we find some unexpected cases of signal-losses. Further technical development could include the characterisation of non-sinusoidal features in spectral baselines, such as with penalised least squares methods (Liu et al. 2022). 

Time-dependent Fourier processing showed some success in suppressing amplitudes of ripples. Mitigation of ripples during data acquisition is likely to be preferable to suppression in post-processing, but such interventions are technically challenging. For sufficiently narrow spectral lines, one practical approach is to allow the target velocity to drift with respect to the telescope, and so distinguish a real feature from a terrestrially-fixed pattern of ripples. 

In a scientific context, the ability to reduce fractional (line-over-continuum) noise down to levels of $\sim10^{-4}$ is very valuable, as a wealth of minor trace gases can become detectable. For example, Palmer et al. (2017) explored complex organics at Titan, detecting vinyl cyanide lines at fractional strengths of a few 10$^{-4}$. These observations were made with the ALMA interferometer, and a future challenge is the suppression of ripple patterns in both the spatial and spectral domains. This will be relevant for future SKA observations of solar system planets -- for example, the detailed exploration of water vapour in Mars' atmosphere (Butler et al. 2004).

\section*{Acknowledgements}

JCMT data were acquired under project ids S16BP007 (PI Greaves) and M22AL006 (PI Clements; data presented here are from dates outside the science runs). The SOFIA data were acquired under project id 75\_0059\_1 (PI Cordiner). The GBT data were acquired under project id AGBT21B-374 (PI Greaves). I thank the respective telescope directors for offering their dedicated time for the first and last of these programs, and the telescope staff for their efforts to make these challenging planetary observations possible. My thanks go also to Dave Clements for a careful reading of the manuscript, and to Sukrit Ranjan whose patient questions about polynomials inspired this work.

%%%%%%%%%%%%%%%%%%%%%%%%%%%%%%%%%%%%%%%%%%%%%%%%%%
\section*{Data Availability}

The data used here are archived at www.cadc-ccda.hia-iha.nrc-cnrc.gc.ca/en/jcmt/ (JCMT), irsa.ipac.caltech.edu/applications/sofia/ (SOFIA) and data.nrao.edu/portal/ (GBT). These archives host data processed to different levels, and the spectra shown here have been post-processed. Post-processed data behind the figures will be uploaded to zenodo.org. The figures were created in SPLAT (http://star-www.dur.ac.uk/~pdraper/splat/splat.html) except for Figure 2 plotted from GAIA (http://star-www.dur.ac.uk/~pdraper/gaia/gaia.html) and Figure 3 plotted from CASA (https://casadocs.readthedocs. io/en/stable/notebooks/image\_visualization.html).



\bibliographystyle{rasti}
\bibliography{example} % if your bibtex file is called example.bib

\noindent Baldwin C., McDermid R.M., Kuntschner H., Maraston C., Conroy C., 2018, MNRAS, 473, 4698 \\
Barnes D.G., Briggs F.H., Calabretta M.R., 2005, Radio Science, 40, RS5S13 \\
Butler B.J., Campbell D.B., de Pater I., Gary D.E., 2004, New Astronomy Reviews, 48, 1511 \\
Cordiner M.A. et al., 2023, GRL, 50, e2023GL106136 \\
Cordiner M.A. et al., 2022, GRL, 49, e2022GL101055 \\
Cram T., 1974, Astronomy and Astrophysics Supplement, 15, 339 \\
de Pater I., Massie S.T., 1985, Icarus, 62, 143 \\
Gordon K.J., Gordon C.P., 1975, Astronomy \& Astrophysics, 40, 27 \\
Greaves J.S., Petkowski J.J., Richards A.M.S., Sousa-Silva C., Seager S., Clements D.L., 2023, GRL, 50, e2023GL103539 \\
Greaves J.S et al., 2022, MNRAS, 514, 2994 \\
Greaves, J.S. et al., 2021, Nature Astronomy, 5, 655. \\
Henkel C., 2005, Effelsberg Memo Series 25022005, at https://eff100mwiki.mpifr-bonn.mpg.de/ \\
Liu B., Wang L., Wang J., Peng B. Wang H., 2022, PASA, 39, e050 \\
Mahieux A. et al., 2023, Icarus, 399, 115556 \\
Matthews H.E., Marten A., Moreno R., Owen T., 2002, ApJ, 580, 598 \\
Palmer M.Y., Cordiner M.A., Nixon C.A., Charnley S.B., Teanby N.A., Kisiel, Z., Irwin P.G.J., Mumma M.J., 2017, Science Advances, 3, e1700022 \\
Richards A.M.S., Moravec E., Etoka S., Fomalont E.B., P\'{e}rez-S\'{a}nchez A.F., Toribio M.C., Laing R.A., 2022, ALMA Memo Series 620, at https://library.nrao.edu/public/memos/alma/ main/memo620.pdf \\
Sandor B.J., Clancy R.T., Moriarty-Schieven G., 2012, Icarus, 217, 839 \\
Starr E., 2024, https://help.almascience.org/kb/articles/what-errors-could-originate-from-the-correlator-spectral-normalization-and-tsys-calibration \\
Walsh A., 2009, https://www.atnf.csiro.au/research/workshops/2009/ astro-informatics/ASAP-tutorials/Tutorial-5.pdf \\

%\begin{thebibliography}{99}
%\end{thebibliography}

%%%%%%%%%%%%%%%%%%%%%%%%%%%%%%%%%%%%%%%%%%%%%%%%%%

%%%%%%%%%%%%%%%%% APPENDICES %%%%%%%%%%%%%%%%%%%%%

% Don't change these lines
\bsp	% typesetting comment
\label{lastpage}
\end{document}